# PERFORMANCE ANALYSIS ON THE BASIS OF A COMPARATIVE STUDY BETWEEN MULTIPATH RAYLEIGH FADING AND AWGN CHANNEL IN THE PRESENCE OF VARIOUS INTERFERENCE

Sutanu Ghosh

Dr. Sudhir Chandra Sur Degree Engineering College

**ABSTRACT**

*Interference is the most important issue for present wireless communication. There are various kinds of channel used in wireless communication. Here I want to show a performance analysis on the basis of two different channels – AWGN and Multipath Rayleigh fading channel. This is the comparative analysis with different kinds of modulation techniques. Here I have also measured the Bit Error Rate with respect to different modulation techniques and compare the rate in different channels. My objective is to compare the different characteristics of the transmitter and receiver for different types of channels and modulators.*

**KEYWORDS**

*Rayleigh fading channel, Bit Error Rate, AWGN, Cellular Interference.*

## I. INTRODUCTION

Rapid Fading is the most important problem for digital mobile communication issues [1]. It mortifies the bit error rate (BER) and repeatedly introduces an exclusive BER. Here, in this issue I want to show the BER modification for different modulation scheme in the different interfering environment.

Channel is the most important subject issue for any kind of communication system. There are various kinds of channel – AWGN channel, Rayleigh Fading channel, Riccian Fading channel [5]. Rayleigh is the well known statistical distribution for amplitude modeling of radio signal in fading environment. Fading is the significant issue for present day wireless communication system. Here I want to describe the comparison between generalized channel AWGN and Rayleigh Fading [2, 3]. Mobile antenna receives a large number of reflected and scattered waves from various directions. Any kind of received signal with Rayleigh distribution instantaneous power follows the exponential distribution property. Thus probability distribution function of power is as follows

$$p(\lambda) = (1/\lambda_0)e^{(-\lambda/\lambda_0)}$$
$$\text{where, } \lambda_0 = E[\lambda] = \int_0^\lambda \lambda\, p(\lambda)\, d\lambda = 2\sigma^2 \ldots\ldots\ldots\ldots\ldots\ldots\ldots\ldots\ldots(a)$$

$E[\lambda]$ is the average value and $2\sigma^2$ is the mean square value.

Due to reflections, scattering, and diffraction of the transmitted signal in a wireless channel, multiple versions of the transmitted signal reach the proper destination at the receiver with different amplitudes and phases. This time dispersive transmitted signal produces flat or frequency selective fading.
I have used flat fading model for this issue. This kind of channel has constant gain and linear phase response over a bandwidth greater than the transmitted signal response [6].

AWGN is the simplest model of a channel. Here I have used the AWGN channel for the comparison purpose. This kind of model is well suited for wired communication. Generally this model is used to take place with well-mannered mathematical models of communication system functionality without fading and distortions.

The fading channel output received signal is given by –

$$r(t) = c(t)s(t) + n(t) \qquad \ldots\ldots\ldots\ldots\ldots(1)$$

where, n(t) is the AWGN with the power spectral density ή in both real and imaginary components. c(t) is the zero mean complex gain of the channel and s(t) is the transmitted signal [4]. c(t) can be formulated as

$$c(t) = \exp(jw_0 t)m(t) \qquad \ldots\ldots\ldots\ldots(2)$$

where, m(t) is the complex Gaussian fading process with variance $\sigma_m^2$.

## II. INTERFERENCE

There are different sources of cellular interference – adjacent channel interference and Co-channel interference. In a cellular communication there are a group of cells which have used same frequency channels, called co-channel cell. The interference come from distant co-channel cell is called as co-channel interference. If the distance between two nearest co –channel cells can be enhanced then the amount of inference can be decreased. It is the part of inter cell interference. Here I have measured the effect of different interfering sources from different cells. The output of all those interfering sources can be added to the original signal and finally they are passed through different kinds of channels. If we want to enhance the capacity of the cellular system, then cell size may be reduced, effectively interference can be increased. The system has required a linear equalizer only develops the spectral characteristics of the interference through its autocorrelation functions [7, 8]. This kind of adaptive equalizer is most essential tool for my result analysis.

## III. SIMULATION MODEL

Here, in Figure 1 I have shown the basic simulation model of my work. It has included a transmitter with binary data generator, signal sources, modulator. A simple raised cosine filter has been used with a roll off factor 0.22. The output of the transmitter can be inserted into the channel with lots of interference. Finally they have come into the receiver. The receiver has included filter, down sampler with down sampling factor 8 and a same type of demodulator used in the transmitter.

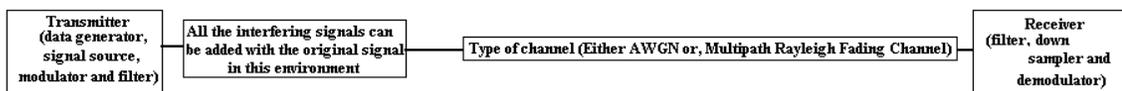

Figure1: Experimental model

## IV. SIMULATION RESULTS

This work is based on the comparison between AWGN channel and Multipath Rayleigh Fading channel. I have considered 3 different modulation techniques – QPSK, 16 PSK AND 64 PSK for this comparative study. I have focused on 3 different matters regarding this issue. Those are transmitted signal response, received constellation and received signal response. Here I have measured and shown all the response on the basis of 30 dB SNR.

Multipath fading has *discrete paths* to propagate the signal in the system. Average path gains of those discrete paths are calculated in decibels. Each of the discrete paths has a certain amount of delay.

Doppler Shift is the parameter provides an estimation of the relative radial velocity of a moving target in multipath channel. Here I have taken an environment that has a Doppler shift of 10 Hz. Figure 2, 3 and 4 shows the different types of transmitting signal response for three different orders of modulation. In my comparison there has a different peak at 0, 2 and -2 MHz for all those types of modulation. I have seen that 64 PSK has the better response on the basis of less amount of interference at center frequency with respect to other types of modulation. But I have examined there have large amount of interference can be experienced for the distant bit interval from center frequency at higher order of modulation.

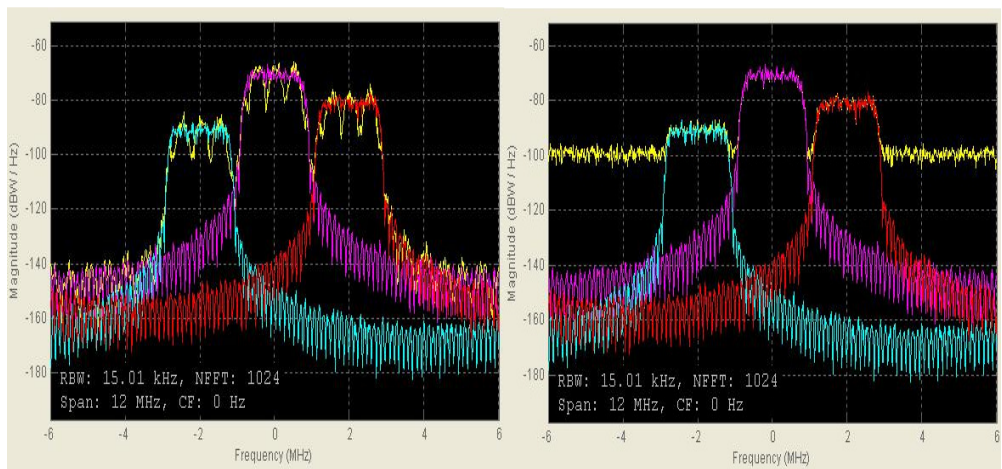

Fig. 2. Comparison between the QPSK transmitted signal response of Multipath Rayleigh fading and AWGN channel

From these results, it's easier to conclude that lower order modulation has higher amount of interference at center frequency with respect to higher order modulation. The external interference is present on the outside of the envelope for AWGN channel. The same interference is present inside the envelope of single bit duration for the Rayleigh fading channel.

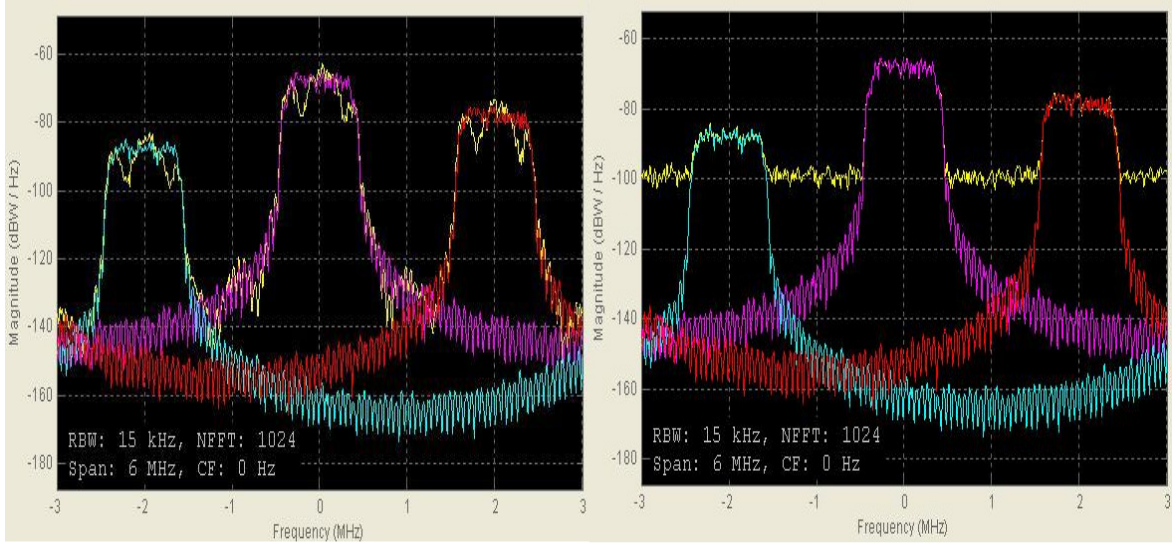

Figure 3: Comparison between the 16 PSK transmitted signal response of Multipath Rayleigh fading and AWGN channel

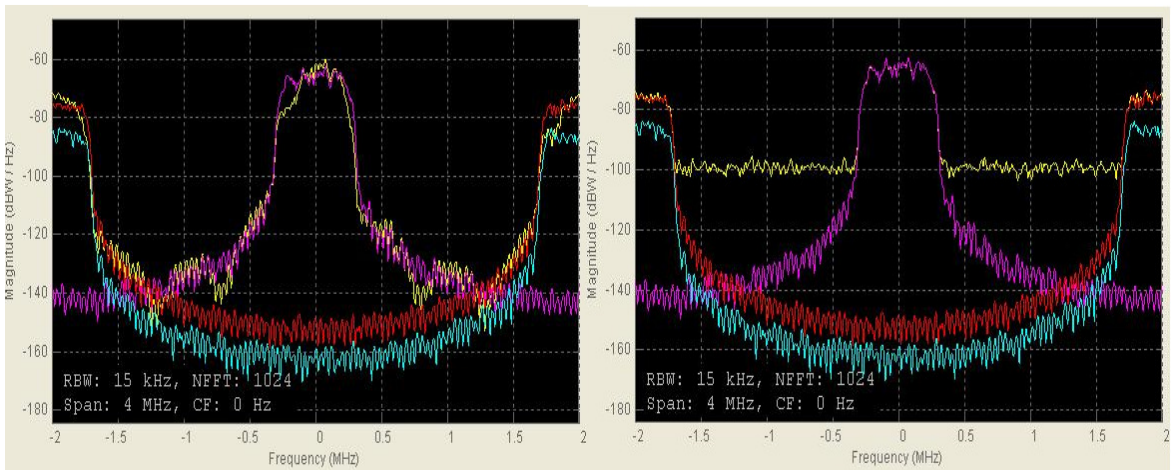

Figure 4: Comparison between the 64 PSK transmitted signal response of Multipath Rayleigh fading and AWGN channel

Figure 5, 6 and 7 depicts the received constellation for different types of channel and modulation. A large number of constellation points can be observed for each and every type of modulation in Multipath Rayleigh fading channel with respect to AWGN channel. If the order of modulation can be increased then the number of constellation points can be enhanced by a multiplicative factor.

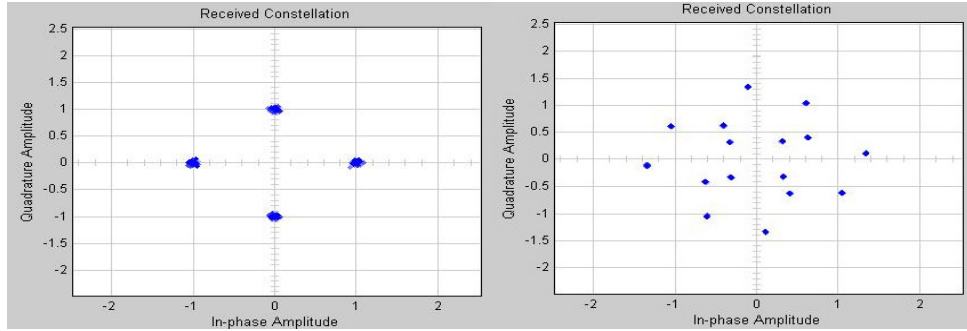

Figure 5: Comparison between AWGN and Multipath Rayleigh fading channel on the basis of received Constellation for QPSK

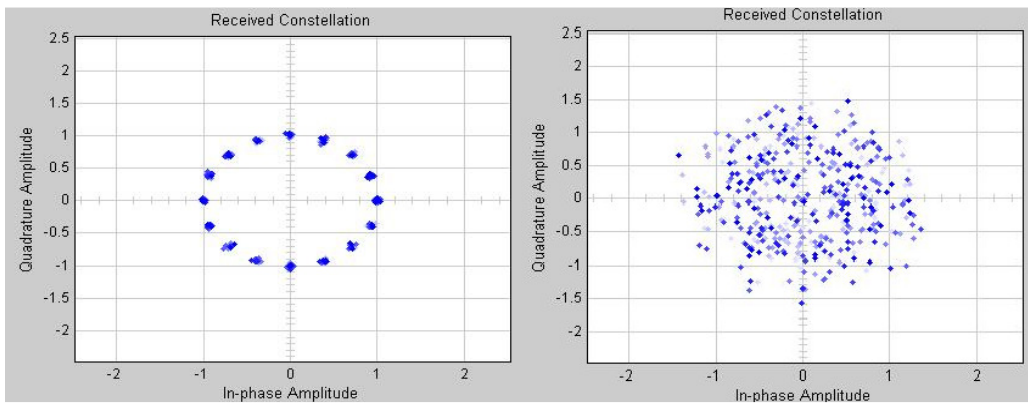

Figure 6: Comparison between AWGN and Multipath Rayleigh fading channel on the basis of received Constellation for 16 PSK

If we compare Figure 6 and 7, then we will see that fig. 6 have random distribution; whereas, Figure 7 have properly arranged distribution.

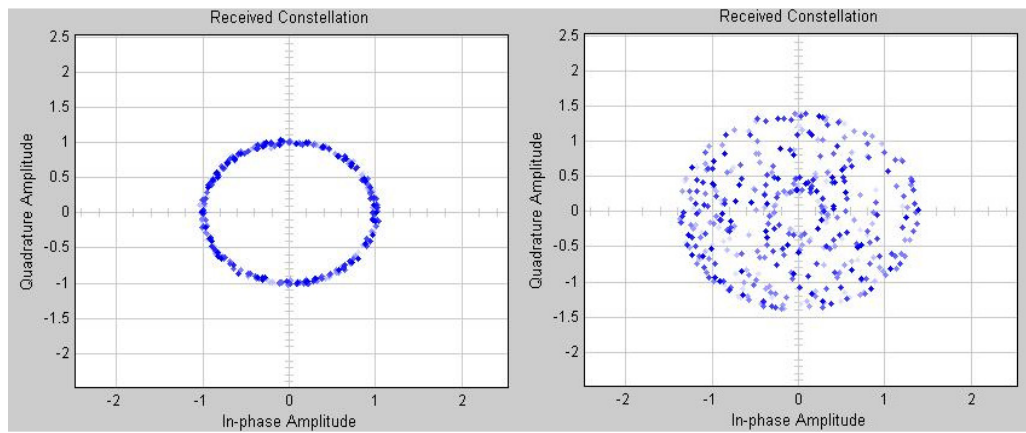

Figure 7: Comparison between AWGN and Multipath Rayleigh fading channel on the basis of received Constellation for 64 PSK

Finally, I have concentrated my work in the receiver section. Figure 8, 9 and 10 shows the received signal response to different kinds of modulation and channel. In case of QPSK, there has a flat response for AWGN channel with respect to other types of modulation. The better response can be observed in 64 PSK modulation for both of the channel.

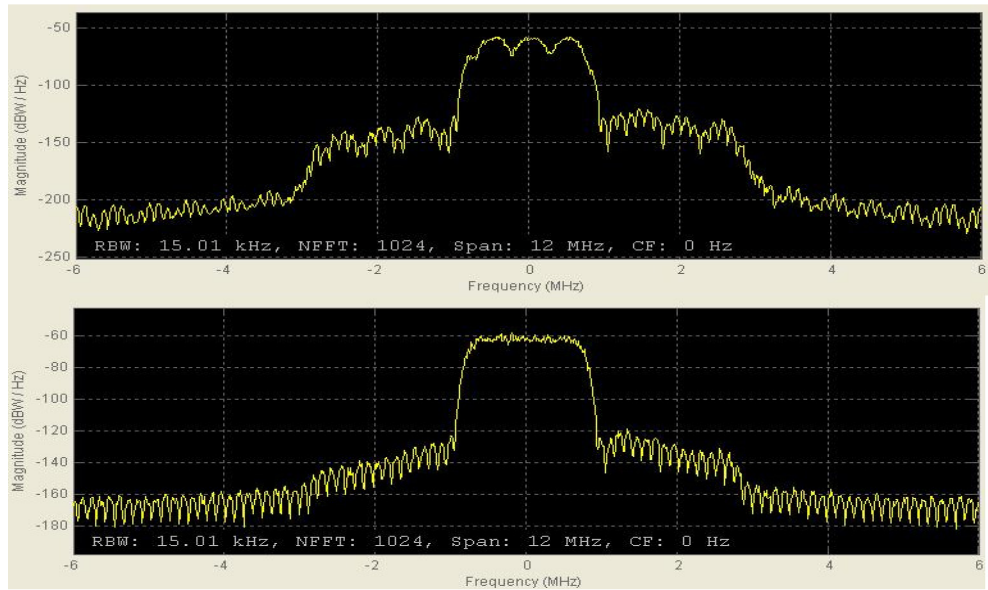

Figure 8: Comparison between the QPSK received signal response of Multipath Rayleigh fading and AWGN channel

From these figures, it is clear that the bit duration of receiving signal for the different types of modulation is much more perfect for multipath Rayleigh fading channel with respect to AWGN channel.

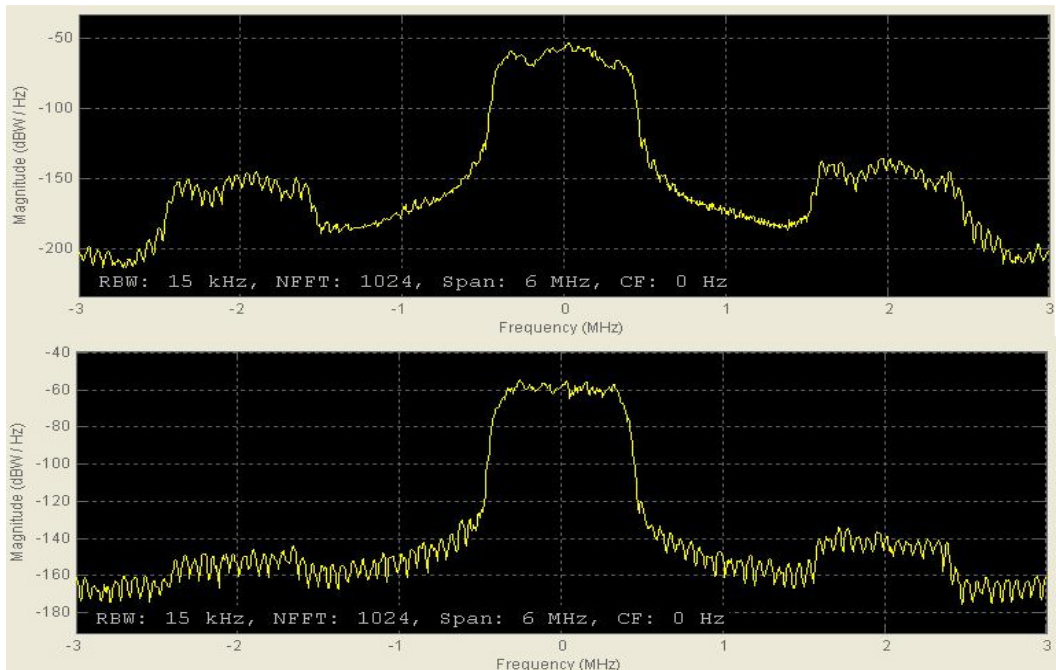

Figure 9: Comparison between the 16 PSK received signal response of Multipath Rayleigh fading and AWGN channel

20

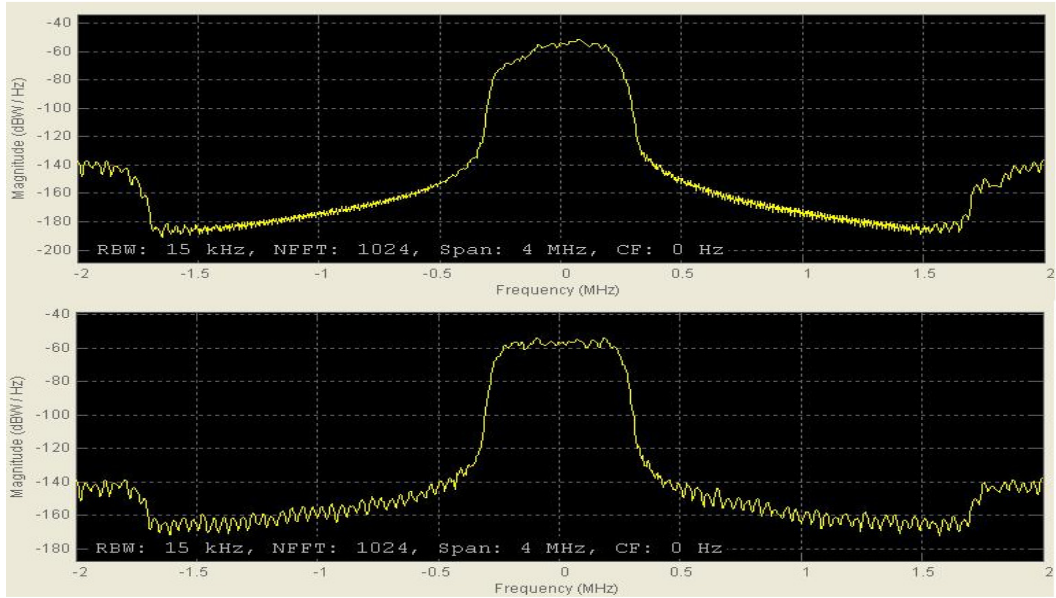

Figure 10: Comparison between the 64 PSK received signal response of Multipath Rayleigh fading and AWGN channel

I have also measured the bit error rate for different channels at different order of modulation. The highest bit error rate can be observed for multipath Rayleigh fading channel at QPSK modulation. The lowest bit error rate can be observed for 16 PSK. From the BER point of view, AWGN channel is much better than multipath Rayleigh fading channel.

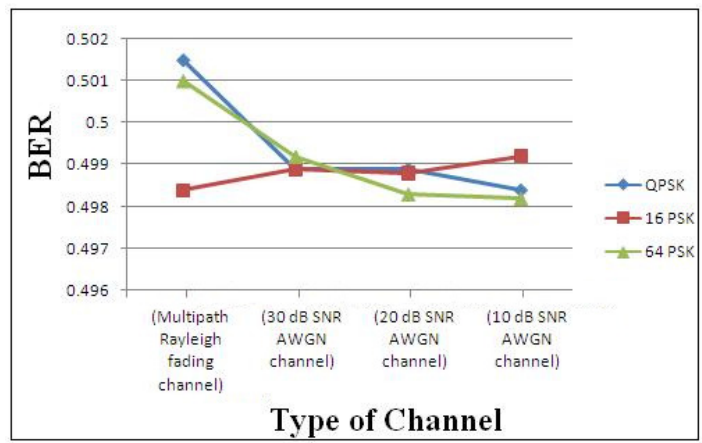

Figure 11: Graphical plot of Bit Error Rate versus the type of channel

Figure 11 gives an idea regarding the bit error rate for different types of modulations at different channels. From this figure it is very clear that the bit error rate is very less for 10 dB SNR AWGN channel with respect to other kinds of channel of communication. In case of 16 PSK, the direction of the movement is opposite with respect to other modulations. That means, it have the bit error rate is less at multipath Rayleigh fading and bit error rate is large for 10 dB SNR AWGN channel. The highest bit error rate 0.5015 can be observed at Rayleigh fading channel for QPSK modulation.

## V. CONCLUSION

For all the mathematical terms and graphical plots it can be concluded that 64 PSK modulation have a better response in multipath Rayleigh fading channel with respect to other kinds modulation. If the amount of Doppler shift frequency can be changed for fading channels then the bit error rate is also changed for each and every type of modulations. Figure 12 gives an exception of this observation at 16 PSK modulation for the Doppler shift of 15 Hz. The interference can be decreased with the increment of up and down sampling factor.

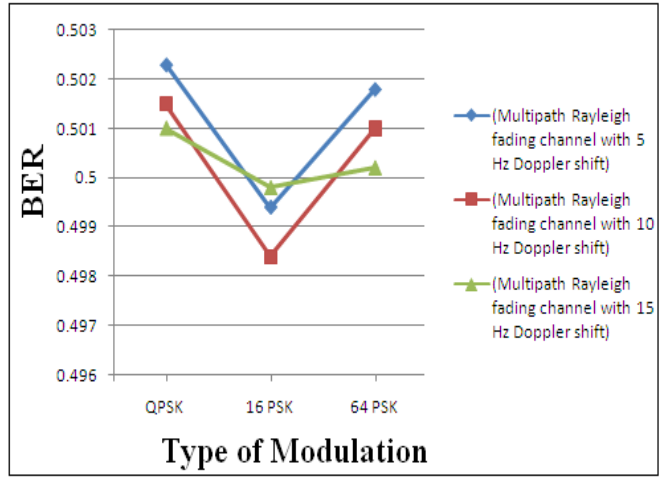

Figure 12: Graphical plot of Bit Error Rate versus type of Modulation for Multipath Rayleigh Fading with different amount of Doppler shift